\documentstyle[editedvolume]{crckapb}
\input{epsf}
\newcommand{\EQ}{\begin{equation}}
\newcommand{\EN}{\end{equation}}
\newcommand{\EQA}{\begin{eqnarray}}
\newcommand{\ENA}{\end{eqnarray}}
\newcommand{\eq}[1]{(\ref{#1})}
\newcommand{\Eq}[1]{Eq.~(\ref{#1})}

\newcommand{\Fig}[1]{Figure~\ref{#1}}

\newcommand{\bra}[1]{\langle #1\rangle}

\newcommand{\meanAA}{\overline{\mbox{\boldmath $A$}}}
\newcommand{\meanBB}{\overline{\mbox{\boldmath $B$}}}
\newcommand{\meanJJ}{\overline{\mbox{\boldmath $J$}}}

\newcommand{\uu}{\mbox{\boldmath $u$} {}}
\newcommand{\vv}{\mbox{\boldmath $v$} {}}

\newcommand{\bb}{\mbox{\boldmath $b$} {}}

\newcommand{\BB}{\mbox{\boldmath $B$} {}}

\newcommand{\AAA}{\mbox{\boldmath $A$} {}}

\newcommand{\jj}{\mbox{\boldmath $j$} {}}
\newcommand{\JJ}{\mbox{\boldmath $J$} {}}

\newcommand{\kk}{\mbox{\boldmath $k$} {}}

\newcommand{\nab}{\mbox{\boldmath $\nabla$} {}}

\newcommand{\oo}{\mbox{\boldmath $\omega$} {}}

\newcommand{\dd}{{\rm d} {}}

\newcommand{\yana}[5]{ (#1) #5. {\em Astron. Astrophys. }{\bf #2}, #3--#4}

\newcommand{\yjfm}[5]{ (#1) #5. {\em J. Fluid Mech. }{\bf #2}, #3--#4}

\newcommand{\yjgr}[5]{ (#1) #5. {\em J. Geophys. Res. }{\bf #2}, #3--#4}
\newcommand{\yapj}[5]{ (#1) #5. {\em Astrophys. J. }{\bf #2}, #3--#4}

\newcommand{\yjour}[6]{ (#1) #6. {\em #2} {\bf #3}, #4--#5}

\newcommand{\ybook}[3]{ (#1) {\em #2}. #3}

\newcommand{\smnS}[2]{  (#1) #2, {\em Mon.\ Not.\ Roy. Astron. Soc.}}

\newcommand{\sapjS}[2]{ (#1)  #2, {\em Astrophys. J.}}

\newcommand{\ea}{{\em et al.\ }}

\begin{opening}
\title{The inverse cascade in turbulent dynamos}
\author{Axel Brandenburg}
\institute{Nordita, Blegdamsvej 17, DK-2100 Copenhagen \O, Denmark\\
Mathematics Department, Univ. of Newcastle, NE1 7RU, UK}
\end{opening}
\begin{document}

\begin{abstract}
The emergence of a large scale magnetic field from randomly forced isotropic
strongly helical flows is discussed in terms of the inverse cascade of
magnetic helicity and the $\alpha$-effect. In simulations of such flows
the maximum field strength exceeds the equipartition field strength for
large scale separation. However, helicity conservation controls the speed
at which this final state is reached. In the presence of open boundaries
magnetic helicity fluxes out of the domain are possible. This reduces
the timescales of the field growth, but it also tends to reduce the
maximum attainable field strength.
\end{abstract}

\section{Introduction}

It was since the mid-seventies when Frisch \ea (1975) and Pouquet \ea
(1976) came up with the idea that the large scale magnetic fields seen
in many astrophysical bodies could be caused by an inverse cascade-type
phenomenon. Although there were close links with earlier results that
helicity and lack of mirror symmetry are important (Steenbeck \ea 1966;
see also Krause \& R\"adler 1980), the notion of an inverse cascade
has put dynamo theory into a self-consistent framework of nonlinear
turbulence theory.

Unfortunately the inverse cascade concept was not easily assimilated
by the astrophysical community. The reasons are simple: the inverse
cascade approach was developed in the framework of isotropic homogeneous
turbulence, and was not readily applicable to astrophysical bodies that
were stratified and enclosed in boundaries. Thus, people continued to
use $\alpha^2$ and $\alpha\Omega$-dynamos (Moffatt 1978), which enabled
modelling of a large variety of astrophysical bodies.

Here we want is to look more closely at inverse cascade dynamos. In
particular, we want to know what kind of field they produce and how
this relates to the fields generated by an $\alpha^2$-dynamo. The full
results of this work are presented in a separate paper (Brandenburg 2000,
hereafter referred to as B2000). In the present paper we also discuss the
effects of open boundaries allowing magnetic helicity fluxes out of the domain
into the exterior or across the equator. This is an important issue that
has been raised recently in the context of dynamo theory (Blackman \&
Field 2000, Kleeorin \ea 2000).

In order to model isotropic random flows we adopt a forcing
function that consists of randomly oriented Beltrami fields with a
wavenumber, $k_{\rm f}$, that is larger than the smallest wavenumber,
$k_1$, that fits into the box. For most of the calculations we use $k_{\rm
f}=5$ and $k_1=1$, leaving some margin for scale separation. In order to see
more clearly the effects of scale separation we also have one run where
$k_{\rm f}=30$. In the wavenumber band $4.5<|\kk|<5.5$ (for $k_{\rm f}=5$)
there are 350 wavevectors which are chosen randomly at each timestep,
so the forcing is $\delta$-correlated in time, but the resulting velocity
field is not. In fact, the velocity has a well defined correlation time
that agrees well with the turnover time $\tau=\ell_{\rm f}/u_{\rm rms}$,
where $\ell_{\rm f}=2\pi/k_{\rm f}$ is the forcing scale and $u_{\rm rms}$
is the rms velocity.

The degree of turbulence that develops depends on the range of length
scales left between the forcing scale and the dissipative cutoff scale. A
reasonable range can only be obtained if the forcing wavenumber is not
too high, so the run with $k_{\rm f}=30$ (Run~6 in B2000) is an example
where the flow is laminar on the forcing scale. The degree of mixing,
as measured by the ratio of the turbulent to the microscopic diffusion
coefficients for a passive scalar, $D_{\rm t}/D$, is here of order one. In
a more turbulent run, Run 3 of B2000, this ratio is around 40. However,
quite independently of how turbulent a run is, we find the emergence
of a large scale magnetic field after some time. The resulting field
resembles closely that obtained from an $\alpha^2$-dynamo with the same
(periodic) boundary conditions. This analogy enables us to make contact
with mean-field theory and to explain the resulting turbulent transport
coefficients.

\section{Emergence of a large scale field}

\begin{figure}[h!]\epsfbox{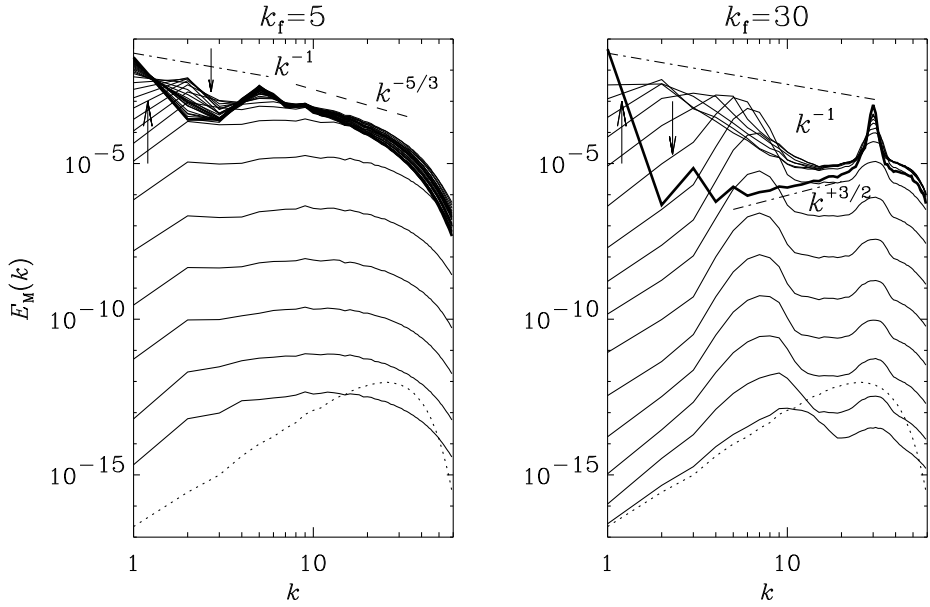}\caption[]{
Magnetic energy spectra for runs with small and large scale separation,
$k_{\rm f}=5$ and 30, respectively.
}\label{pspec_growth_both}\end{figure}

The inverse cascade is traditionally described in terms of energy spectra.
In \Fig{pspec_growth_both} we compare the spectral field evolution
for two different forcing wavenumbers. Note that the envelope of the
magnetic energy fits underneath a $k^{-1}$ slope. The peaks at $k=k_1$
and $k_{\rm f}$ also fit underneath the same slope. There are several
features of the spectrum that are characteristic also of other cases
investigated. For large enough scale separation one sees that the
magnetic energy grows fast at two distinct wavenumbers, $k\approx30$ and
$k\approx7$. However, when the energy at $k\sim7$ reaches saturation the
energy begins to be transferred to larger scales until much of the
magnetic energy is at the largest scale possible. During this phase
the magnetic energy at intermediate scales decreases to some minimum
value which follows roughly a $k^{+3/2}$ spectrum. This effect may be
referred to as `self-cleaning', because by removing energy at intermediate
scales the field at the largest scales appears less perturbed and hence
cleaner. This self-cleaning effect is the result of nonlinearity, which
suppresses the growth at intermediate scales. However,
the type of nonlinearity does not seem to matter: even with ambipolar
diffusion the same behaviour is found (Brandenburg \& Subramanian 2000).

\epsfxsize=12.6cm

\begin{figure}[h!]\epsfbox{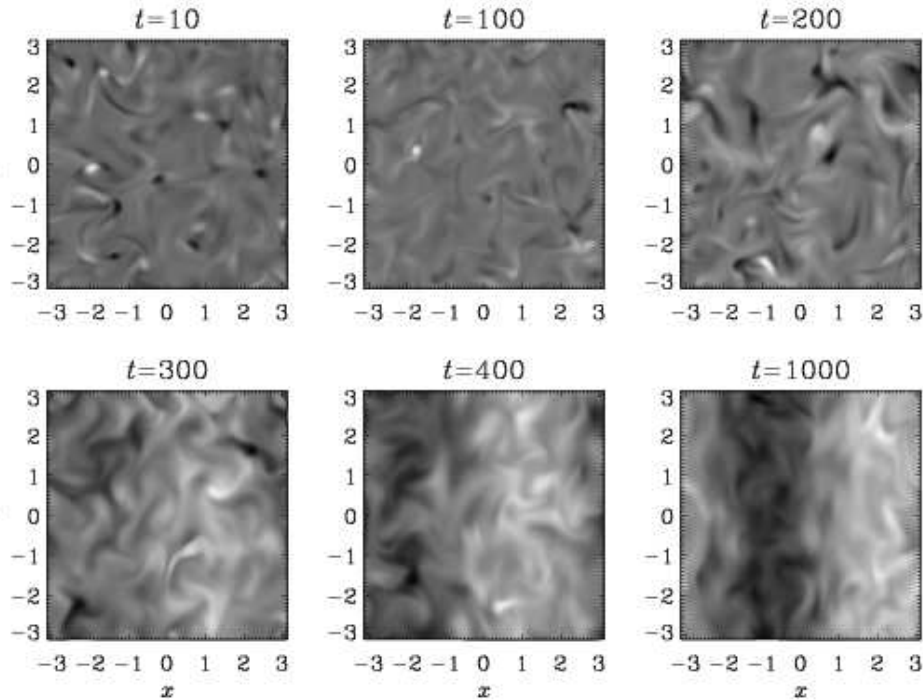}\caption[]{
Gray-scale images of cross-sections of $B_x(x,y,0)$ for Run~3 of
B2000 at different times showing the gradual build-up of
the large scale magnetic field after $t=300$. Dark (light) corresponds
to negative (positive) values. Each image is scaled with respect to its
min and max values.
}\label{Fpimages_hor}\end{figure}

In \Fig{Fpimages_hor} we show cross-sections of $B_x$
at different times. Towards the end of the evolution the large scale
magnetic field is essentially a Beltrami field which is here of the form
$\meanBB\sim(\sin y,0,\cos y)$, apart from some phase shift in $y$.

\epsfxsize=12.6cm

\section{The final equilibrium field strength}

\begin{figure}[h!]\epsfbox{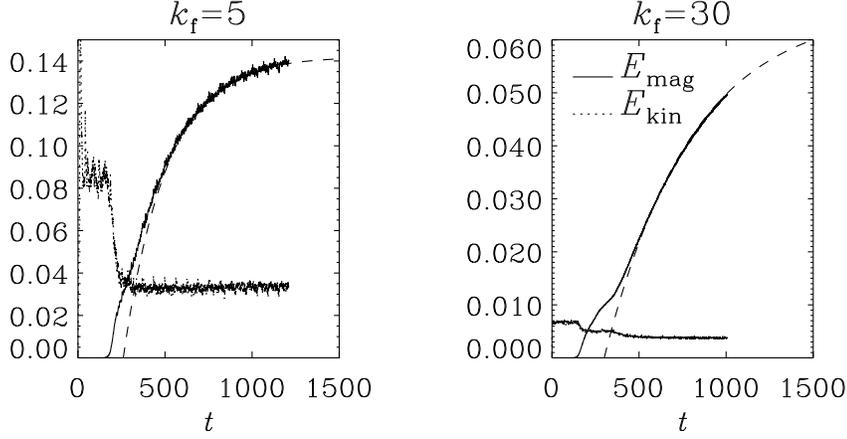}\caption[]{
Evolution of kinetic (dotted) and magnetic (solid) energies. The slow
evolution of the magnetic energy follows approximately the $1-e^{-2\eta
\Delta t}$ behaviour (dashed line) of \Eq{Bconstr} that results from
helicity conservation.
}\label{Fpn_comp}\end{figure}

Characteristic to all the runs reported in B2000 is the fact that
super-equipartition field strengths are reached. In \Fig{Fpn_comp}
the evolution of magnetic and kinetic energies are shown for two
runs with different forcing wavenumbers. Note also the extremely slow
evolution past the moment where the kinetic energy drops suddenly to a
smaller value. This is when saturation of small scale magnetic energy
is reached. However, after that moment the large scale magnetic energy
continues to grow for some time, because the resulting large scale field
is force-free and does hence not affect the velocity field directly.

\epsfxsize=12.6cm

The prolonged saturation behaviour found in the present simulations
is at first glance unusual. Since at late times most of the magnetic
energy is in the large scales, this slow evolution must have to do
with the properties of the large scale field. An important property of
this large scale field is that it possesses magnetic helicity. At the
same time magnetic helicity is conserved by the nonlinear terms and can
hence only change resistively on an ohmic timescale. (The case with open
boundaries is different and will be discussed separately.) In order to
demonstrate magnetic helicity conservation we consider a periodic box
and write the magnetic field, $\BB$, as the curl of a vector potential,
$\AAA$, so $\BB=\nab\cdot\AAA$. We use the uncurled induction equation,
\EQ
\partial\AAA/\partial t=\uu\times\BB-\eta\JJ-\nab\phi,
\EN
where $\JJ=\nab\times\BB$ is the current density, $\eta$ is the magnetic
diffusivity, and $\phi$ is the electrostatic potential. The magnetic helicity,
which can be defined as $\bra{\AAA\cdot\BB}$, satisfies
\EQ
\dd\bra{\AAA\cdot\BB}/\dd t=-2\eta\bra{\JJ\cdot\BB},
\label{helcons}
\EN
where angular brackets denote volume averages over the periodic
domain. The steady state solution must satisfy $\bra{\JJ\cdot\BB}=0$,
so small and large scale current helicities must be comparable in
magnitude, $|\bra{\jj\cdot\bb}|\approx|\bra{\meanJJ\cdot\meanBB}|$,
but of opposite sign. The magnetic helicity is however concentrated on
large scales (its spectrum is $k^2$ times that of the current helicity),
so its small scale contribution is negligible.

We measure the degree of magnetic helicity of the large scale field by the
quantity $k_{AB}^{-1}=|\bra{\meanAA\cdot\meanBB}|/\bra{\meanBB^2}$, which is
a length scale characterizing the scale of the helical contribution. In
a periodic domain of size $L$ the smallest wavenumber is $k_1=2\pi/L$,
and so $k_{AB}$ is bounded from above by $k_{AB}\leq k_1$. The large
scale current helicity is $k_1^2$ times the magnetic helicity, so we have
\EQ
\bra{\meanAA\cdot\meanBB}=\mp\bra{\meanBB^2}/k_{AB}=
\bra{\meanJJ\cdot\meanBB}/k_1^2,
\EN
where the upper sign applies to the case of positive
kinetic helicity in the turbulence. Using this in \Eq{helcons} together
with $\bra{\AAA\cdot\BB}\approx\bra{\meanAA\cdot\meanBB}$ and
$\bra{\JJ\cdot\BB}=\bra{\meanJJ\cdot\meanBB}+\bra{\jj\cdot\bb}$ yields
\EQ
\mp\dd\bra{\meanBB^2}/\dd t=\pm2\eta k_1^2\bra{\meanBB^2}
-2\eta k_{AB}\bra{\jj\cdot\bb}.
\label{Bconstr_ODE}
\EN
Prior to saturation $\bra{\jj\cdot\bb}$ is small, but during saturation
its value is limited by the kinetic helicity, so
\EQ
\bra{\jj\cdot\bb}\approx\bra{\rho}\bra{\oo\cdot\uu}
\approx\pm k_{\rm f}\bra{\rho\uu^2}
=\pm k_{\rm f} B_{\rm eq}^2/\mu_0.
\EN
This yields an evolution equation for the mean magnetic field,
\EQ
{\bra{\meanBB^2}\over B_{\rm eq}^2}\approx
{k_{\rm f}k_{AB}\over k_1^2}\left\{1-
\exp\left[-2\eta k_1^2 (t-t_{\rm sat})\right]\right\},
\label{Bconstr}
\EN
which is only valid at late times when $\Delta t\equiv t-t_{\rm sat}>0$.
(Here, $t_{\rm sat}$ is the time when the small scale field saturates.)
We emphasize that this relation is rather general and independent of
the actual model of field amplification, because we used
only the concept of magnetic helicity conservation. The important
point here is that full saturation is only obtained after an ohmic
diffusion time. In that sense \Eq{Bconstr} poses a constraint on the mean
magnetic field at late times. It applies only as long as the magnetic
field is helical. Indeed, lower degrees of helicity, i.e.\ smaller values
of $k_{AB}^{-1}$, allow larger values of the final field strength; see
\Eq{Bconstr}. This has been verified in a model where differential
rotation or shear contributed significantly to the field amplification
(Brandenburg \ea 2000). However, this only relaxes the constraint by
a certain factor which depends on the degree of helicity of the large
scale field which, in turn, depends on the degree of linkage of poloidal
and toroidal field. We found that to a good approximation this factor is
given by the ratio $Q$ of toroidal to poloidal field strengths. In the sun
this factor is less than a hundred, so this is a relatively minor effect
compared with the value of the magnetic Reynolds number ($10^8-10^{10}$).

\section{Helicity exchange across the equator}

\begin{figure}[h!]\epsfbox{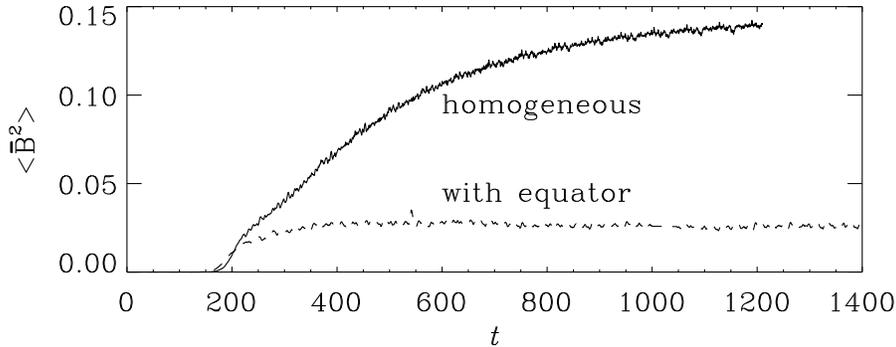}\caption[]{
Evolution of the magnetic energy for a run with homogeneous forcing
function (solid line) and a forcing function whose helicity varies
sinusoidally throughout the domain (dotted line) simulating the effects
of an equators.
}\label{Fpbmean_Run2}\end{figure}

A different way of relaxing the slow growth problem is to allow for
fluxes out of the dynamo volume either into the exterior of the dynamo
(the corona or halo) or from one hemisphere to the other. In any case,
there would be an extra surface term in \Eq{helcons}. Here we want
to discuss the latter alternative of a helicity flux between the two
hemispheres. Such a flux could result from a turbulent exchange of
magnetic helicity between the two hemispheres and should therefore
be proportional to some turbulent magnetic diffusivity $\eta_{\rm
t}$. Based on dimensional arguments, one may expect such a term to be of
the form $\eta_{\rm t}k_{\rm eff}^2H$, where $H$ is the gauge-invariant
magnetic helicity for open volumes (Berger \& Field 1984), which replaces
$\bra{\AAA\cdot\BB}$. The diffusion of magnetic helicity depends on the
length scale $2\pi/k_{\rm eff}$ over which the magnetic helicity varies
(if evaluated over different volumes), so we expect $k_{\rm eff}\leq
k_1$. Equation \eq{Bconstr_ODE} becomes then
\EQ
\mp\dd\bra{\meanBB^2}/\dd t=
\pm2\eta k_1^2\bra{\meanBB^2}
-2\eta k_{AB}^{-1}\bra{\jj\cdot\bb}
\pm2\eta_{\rm t} k_{\rm eff}^2\bra{\meanBB^2},
\EN
so the solution for the mean magnetic field is (assuming again
$\bra{\jj\cdot\bb}\approx\pm k_{\rm f}B_{\rm eq}^2/\mu_0$) given by
\EQ
{\bra{\meanBB^2}\over B_{\rm eq}^2}\approx
{\eta k_{\rm f}k_{AB}\over\eta k_1^2+\eta_{\rm t}k_{\rm eff}^2}
\left\{1-\exp\left[-2\left(\eta k_1^2+\eta_{\rm t} k_{\rm eff}^2\right)
\left(t-t_{\rm sat}\right)\right]\right\}.
\label{Bconstr_wflux}
\EN
Thus, the time dependence is no longer resistively dominated, because
the microscopic diffusivity is now supplemented by an additional turbulent
diffusivity. Unfortunately, however, the amplitude of the final field
decreases in such a way that the initial linear growth in unchanged.
This is simply because of the fact that the flux term, as modelled here,
does not act as an effective driver, which is what the $\bra{\jj\cdot\bb}$-term
did. This is also clearly seen in a simulation where we have included an
equator by modulating the forcing function such that the kinetic helicity
varied sinusoidally n the $z$-direction within the domain; see
\Fig{Fpbmean_Run2}.

\epsfxsize=12.6cm

\section{Conclusions}

The present work has shown that for helical velocity fields a large scale
magnetic field is generated. There are strong parallels with the fields
resulting from mean-field $\alpha^2$-dynamos (see B2000 for details). One
aspect that has now begun to receive major attention is related to magnetic
helicity conservation, which prevents rapid growth of magnetic helicity
and hence helical large scale magnetic fields. Shear, which corresponds to
differential rotation in a rotating system, relaxes this constraint only
partially in that it lowers the fraction of the field that contributes
to magnetic helicity. On the other hand, by allowing magnetic flux to escape
through the boundaries, or allowing for mixing of magnetic helicity of
opposite sign at the equator, the helicity constraint is modified such that
the time scale is no longer resistively dominated. The problem however
is that various attempts to model this effect result in significantly 
lower equilibrium amplitudes of the magnetic field. Part of the problem
is that the loss of magnetic helicity implies at the same time a loss
of magnetic energy. It  would therefore be advantageous for the dynamo
to lose preferentially small scale magnetic helicity and energy. This is
something that the dynamo in the computer keep refusing to do. It may
therefore be important to resort to more realistic simulations where the
flows are driven naturally and not by some artificial stirring in space.

\end{document}